\documentclass[%
 reprint,
 nofootinbib,
 amsmath,amssymb,
 aps,
]{revtex4-2}

\usepackage{amssymb}
\usepackage{graphicx}
\usepackage{dcolumn}
\usepackage{bm}
\usepackage[colorlinks=true,citecolor=blue,urlcolor=blue,linkcolor=red]{hyperref}
\usepackage{physics}
\usepackage{siunitx}
\usepackage{pgf}
\usepackage[normalem]{ulem}
\usepackage{xspace}
\usepackage{makecell}

\newcommand{\lcdm}{\mbox{$\Lambda$CDM}}
\newcommand{\mnu}{\mbox{$\Sigma m_\nu$}}

\newcommand{\sptthreeg}{SPT-3G}
\newcommand\rs{r_{\rm d}}
\newcommand\thetas{\theta_s^{\star}}

\newcommand\omm{\omega_{\rm m}}
\newcommand\omddm{\omega_{\rm ddm}^{\rm ini}}
\newcommand\omcb{\omega_{\rm cb}}
\newcommand\omnu{\omega_{\nu}}
\newcommand\mnue{\bar{m}_{\nu}}
\newcommand\omnue{\omega_{\bar{m}_\nu}}

\newcommand{\jcap}{JCAP}

\makeatletter
\newcommand\thefontsize[1]{{#1 The current font size is: \f@size pt\par}}
\makeatother

\begin{document}

\title{What's the matter with $\mnu$?}

\author{Gabriel P. Lynch$^1$ }
\email{gplynch@ucdavis.edu}
\author{Lloyd Knox$^1$}

\affiliation{
$^1$Department of Physics and Astronomy, University of California, Davis, CA, USA 95616
}

\date{\today}

\begin{abstract}
Due to non-zero neutrino rest masses we expect the energy density today in non-relativistic matter, $\omm$, to be greater than the sum of baryon and cold dark matter densities, $\omcb$. We also expect the amplitude of deflections of CMB photons due to gravitational lensing to be suppressed relative to expectations assuming massless neutrinos. The combination of CMB and BAO data, however, appear to be defying both of these expectations. Here we review how the neutrino rest mass is determined from cosmological observations, and emphasize the complementary roles played by BAO and lensing data in this process. We then use a phenomenological model to find that the preference from CMB and BAO data for a matter density that is below expectations from the CMB alone is at the $2.3\, \sigma$ level. We also show that if a fraction of the dark matter decays to dark radiation, the preference for $\omm > \omcb$ can be restored, but with a small increase to the CMB lensing excess. 

\end{abstract}

\maketitle

\section{\label{sec:intro}Introduction}

The standard Big Bang model predicts a sea of relic neutrinos that decoupled at $T \sim 1\text{ MeV}$. The contribution of this neutrino background to the overall energy density has been detected using precise measurements of the cosmic microwave background (CMB), and is consistent with the presence of three massless neutrino species \citep{Planck:2018vyg, SPT-3G:2024atg, ACT:2025tim, SPT-3G:2025bzu}. Neutrinos are known to have mass, however, from atmospheric and solar neutrino oscillation observations, with a sum of neutrino masses $\mnu$ that must exceed about 0.06 eV \citep{ParticleDataGroup:2022pth}. Current upper limits from cosmological observations are approaching this bound \citep{Tristram:2023haj, ACT:2023kun, DESI:2024mwx, SPT-3G:2024atg, SPT-3G:2025zuh, SPT-3G:2025bzu}. A clear detection of $\mnu$ from cosmological observations may occur in the near future and would constitute a milestone in both cosmology and particle physics \citep{Green:2021xzn}. 

However, not only has such a detection not yet occurred, but two different features of the data have emerged which are surprisingly opposite to what one would expect as signatures of neutrino mass. 
The first is that the reconstructed lensing power from Planck \citep{carron_cmb_2022}, the Atacama Cosmology Telescope (ACT) \citep{ACT:2023dou}, and \sptthreeg\ \citep{SPT-3G:2024atg} is higher than expectations from Planck primary CMB and DESI baryon acoustic oscillation (BAO) data, assuming \lcdm. As demonstrated in \citet{SPT-3G:2024atg}, this preference for excess lensing power holds regardless of which Planck likelihood is used, and regardless of which lensing reconstruction is used. By artificially scaling the lensing amplitude with a parameter $A_{\rm lens}$\footnote{Multiple definitions of lensing consistency parameters exist in the literature. Here we adopt the convention of \citet{SPT-3G:2024atg}: $A_{2 \mathrm{pt}}$ scales the amplitude of the lensing potential used to lens the primary CMB, $A_{\rm recon}$ scales the amplitude of  model $C_{\ell}^{\phi \phi}$ used to fit the lensing reconstruction, and when $A_{2\mathrm{pt}} = A_{\rm recon}$ is enforced we denote the combined parameter as $A_{\rm lens}$. With this convention, the Planck $A_L$ is equivalent to $A_{2 \mathrm{pt}}$. If the lensing reconstruction is not included, $A_{2 \mathrm{pt}}$ is functionally equivalent to $A_{\rm lens}$, and $A_{\rm recon}$ would be totally unconstrained.}, 
they found that the combination of Planck primary CMB, BAO, and all three of the aforementioned lensing reconstructions favors \citep{SPT-3G:2024atg}
\begin{equation}
    A_{\rm lens} = 1.083 \pm 0.032,
\end{equation}
a $2.7\, \sigma$ detection of excess lensing power.

Another phenomenological model for quantifying this excess lensing power was introduced earlier in 2024 by  \citep{Craig:2024tky} and subsequently studied in \citep{Green:2024xbb}. Instead of $A_{\rm lens}$ they introduced a neutrino-mass-like parameter, $\Sigma \tilde m_\nu$, that controls a template for adjusting the model CMB lensing power. For negative values of $\Sigma \tilde m_\nu$ lensing power is enhanced, to the same degree as the suppression caused by $\mnu = |\Sigma \tilde m_\nu |$. They found $\Sigma \tilde m_\nu < 0$ at $2.4\, \sigma$. Other signed neutrino mass parameterizations have found consistent results \citep{Elbers:2024sha}.

In this paper, we discuss and quantify the second line of evidence. Allowing for the possibility that the comoving density of non-relativistic matter can decrease after recombination, we show that this scenario is in fact favored by CMB and BAO data. This is also the opposite of what is expected from massive neutrinos: since they become non-relativistic after recombination, we expect them to contribute to the total physical matter density today, $\omm$. That the interaction of BAO and CMB constraints on the background evolution prefers $\omm < \omcb$ (where $\omcb$ is the density of baryons and cold dark matter) was explained by \citet{Loverde:2024nfi}, but not quantified. We quantify this preference here by phenomenologically extending the effect of neutrino masses on the background expansion to negative values. Using Planck PR3 primary CMB data and BAO data from DESI DR2 \citep{DESI:2025zgx}, we find a $2.3\, \sigma$ preference for $\omm$ being less than expected from CMB data alone (assuming $\mnu=0.06\, {\rm eV}$). We refer to this as a ``matter-density deficit."

We also consider the complementary roles played by CMB lensing data and BAO data regarding constraints on neutrino mass. Understanding how constraints from these data sets interact is useful in clarifying how each contribute to current bounds on the neutrino mass. We describe how the sensitivity of CMB lensing data to $\mnu$ relates to the angular scales being probed, and explain why BAO and CMB lensing are powerful probes of $\mnu$ when combined.

The origin of these observational signatures (the matter-density deficit, and the CMB lensing power excess) is unclear, as is whether or not they share a common origin. It is possible that a single model could explain both features, or that they are resolved separately. We work out an example, involving decaying dark matter along with a free neutrino mass, that explains the $\omm < \omcb$ preference, but without addressing the excess lensing problem. Until these features of the data are better understood, current tight limits, derived under the assumption of the $\lcdm + \mnu$ model, should be interpreted with care. 

This paper is organized as follows. In Section~\ref{sec:theory} we review how CMB and BAO data can constrain neutrino masses, both separately and in conjunction. In Section~\ref{sec:background_analysis} we introduce our signed neutrino mass model, discuss our analysis methodology, and quantify the statistical significance of the matter-density deficit. We then introduce a DDM model in Section~\ref{sec:DDM} which is similar to the signed neutrino mass model at the background level, and discuss the constraints in this model space. We briefly review possible solutions to these problems in Section~\ref{sec:discussion}, before concluding in Section~\ref{sec:conclusion}.

\section{Review of constraints on the neutrino mass}\label{sec:theory}

With this section we aim to provide the reader with a conceptual understanding of how cosmological observations, CMB and BAO in particular, enable constraints on the neutrino mass. These observables are sensitive to the gravitational influence of massive neutrinos, a sensitivity which fundamentally makes these constraints possible. For neutrinos in mass ranges of interest, which become non-relativistic after recombination, a combination of probes is necessary to isolate this influence. 

The measurement of $\mnu$ from cosmological data can be thought of as proceeding in two steps. First, the CMB temperature and polarization power spectra are used to estimate the baryon and cold dark matter densities, along with the angular size of the sound horizon; together, these give a constraint on the distance to the last-scattering surface. Knowing this distance is not sufficient to determine $\mnu$, as we can simultaneously vary the cosmological constant and $\mnu$ to keep that distance fixed. To break the degeneracy we can then either add BAO data, which (in combination with the primary CMB data) constrain $\omm = \omcb + \omnu$, or CMB lensing data, which constrain a different linear combination that is approximately $\omcb - 0.7\omnu$. We explain these steps in detail in the remainder of this section.

\subsection{Primary CMB}

First we discuss how the mass/energy densities in the various components of the \lcdm\ model can be determined from CMB data if one assumes the \lcdm\ model, before moving on to the case of the \lcdm\ model extended to include an unknown spectrum of neutrino masses.

\subsubsection{Component densities determined assuming \lcdm\ and ignoring CMB lensing}

In \lcdm\ the energy density in relativistic particles (radiation density) is completely determined by the FIRAS determination of the CMB temperature \citep{Fixsen:1996nj} and assumptions relevant to the thermal production of three species of neutrinos. To be concrete, for now we will assume three species of massless neutrino, so that all of their energy is included in the radiation density. The remainder of the mass/energy density is controlled by three parameters which we can take to be the energy densities today of baryons, $\omega_{\rm b}$, and cold dark matter, $\omega_{\rm c}$\footnote{A mass density specified with the parameter $\omega$ is the mass density in units of $3\times (100 \rm{km/sec/Mpc})^2/(8\pi G) \simeq 1.88 \times 10^{-29} {\rm g/cm}^3$.}, and the angular size of the sound horizon at the epoch of recombination, $\thetas$. The other \lcdm\ parameters describe the spectrum of primordial fluctuations ($A_s$ and $n_s$) and the reionization history, $\tau_{\rm reio}$. 

All three of these quantities ($\omega_{\rm b}$, $\omega_{\rm c}$, $\theta_{\rm s}^*$) can be determined with a high degree of precision from CMB temperature anisotropies, polarization anisotropies, or the combination with very little (although some, as we will explore) sensitivity to potential departures from \lcdm\ in the post-recombination universe. For a discussion of the physics that allows one to infer $\omega_{\rm c}$, $\omega_{\rm b}$, and $\thetas$ (as well as the other three \lcdm\ parameters) from CMB anisotropy data, see \citet{Planck:2016tof}. 

Obviously $\thetas$ is not a density, but given the other two and the \lcdm\ model, the dark energy density can be inferred, as we now explain. The angular size of the sound horizon is the angle subtended by the sound horizon, projected from the last-scattering surface. It is given by $\thetas = r_s^{\star} / D_A^\star$ where $r_s^{\star}$ is the comoving size of the sound horizon at last scattering and $D_A^\star$ is the comoving angular-diameter distance to last scattering. These depend on the expansion rate and sound speed in the plasma via:
\begin{equation}\label{eqn:rs_da}
    r_s^{\star} = \int_{z_\star}^\infty \frac{c_s(z)}{H(z)} \dd z \qand D_A^{\star} = \int_0^{z_\star} \frac{c}{H(z)} \dd z
\end{equation}
where $z_\star$ is the redshift of recombination, which for the standard recombination scenario is around $z_\star \approx 1090$. The expansion rate prior to recombination depends only on the baryon, cold dark matter, and radiation densities, as the cosmological constant contribution is negligibly small. Additionally, of the \lcdm\ parameters, $\omega_{\rm b}$ and $\omega_{\rm c}$ are all that is needed to determine $z_\star$ and $r_s^\star$. Therefore, given $\thetas$, $\omega_{\rm b}$, and $\omega_{\rm c}$ one can determine $D_A^\star = r_s^\star / \thetas$. The only unknown density in the integrand for calculating $D_A^\star$ is $\omega_\Lambda$, and its value can be fixed by matching the previously determined value of $D_A^\star$. With the value of $\omega_\Lambda$ set in this way, $H(z)$ is also known at all redshifts. See \cite[e.g.][]{Knox:2019rjx} for more details.

\subsubsection{Component densities determined assuming \lcdm\ + free $\mnu$}

In the previous section we modeled neutrinos as massless particles; however, from observations of solar and atmospheric neutrinos it is known that at least two neutrino species have non-zero mass. Current global analyses of neutrino oscillation data have constrained the squared mass differences to be $\Delta m^2_{21} \approx 7.5 \times 10^{-5} {\rm \ eV}^2$ and $|\Delta m^2_{31}|\approx 2.55 \times 10^{-3}{\rm\ eV}^2$ \citep{deSalas:2020pgw, Esteban:2020cvm}. Depending on the sign of $\Delta m^2_{31}$, these constraints set lower bounds for the sum of the masses: in the normal ordering (i.e. $m_3 > m_1$, hereafter NO) we have $\mnu>0.06 {\rm\ eV}$, and in the inverted ordering (IO), $\mnu>0.1 {\rm\ eV}$. Because near-term cosmological data will not measure individual masses \citep{Archidiacono:2020dvx}, the cosmological impact of neutrino mass is usually parameterized by the sum of their masses, $\mnu$.\footnote{Throughout this paper we assume two species of massless neutrino and one massive, unless otherwise stated.}

\begin{figure}[t!]
    \centering
    \includegraphics{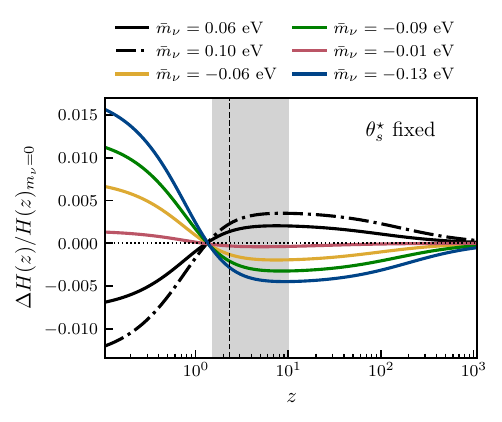}
    \caption{Similar to Fig.~1 of \citet{Pan:2015bgi} we show here fractional changes to the expansion rate at fixed angular size of the sound horizon, as projected from the last-scattering surface. The baseline expansion rate is for a model with $\mnu = 0$. The $\mnue$ parameter and the associated phenomenological model are described in Section~\ref{sec:model} — positive values of $\mnue$ are identical to physical neutrinos with $\mnu = \mnue$ at the background level. The gray band indicates the full width at half maximum of the CMB lensing kernel, showing the redshifts to which CMB lensing is most sensitive. The vertical dashed line is at $z=2.33$, the maximum redshift for which we include BAO measurements.}
\label{fig:Hubble_response} 
\end{figure}

Freeing $\mnu$ changes the picture described previously in two ways. First, the post-recombination total energy density, and thus expansion rate, depends both on $\omega_\Lambda$ and on the unknown neutrino density. Given just $\thetas$, $\omega_{\rm b}$, and $\omega_{\rm c}$ there is now a degeneracy between $\omega_\Lambda$ and $\mnu$, so $H(z)$ is no longer completely determined. Instead, we have a one-parameter family of $H(z)$ curves which all give the same $D_A^\star$ and therefore $\thetas$. Some members of this family are shown in Fig.~\ref{fig:Hubble_response}.

A second way neutrino masses can alter the picture presented so far arises when neutrinos become non-relativistic around or before recombination. Defining the non-relativistic transition to occur when the rest mass is equal to the average kinetic energy we get
\begin{equation}
\label{eqn:znr}
    1+ z_{\rm nr} = \frac{m_\nu}{3.15 T_{\nu0}} \approx 113 \frac{m_{\nu}}{0.06 \text{ eV}}.
\end{equation}
This transition occurs near recombination or earlier for neutrinos more massive than $m_{\nu} \sim 0.6\ {\rm eV}$. Masses in this range will lead to changes in early ISW effects, providing a new CMB signal (beyond what is captured by $\omega_{\rm b}$, $\omega_{\rm c}$, and $\thetas$ inferences) and allowing for the degeneracy between $\omega_\Lambda$ and $\Sigma m_\nu$ to be partially broken \cite{Hou:2012xq, Lesgourgues_Mangano_Miele_Pastor_2013}. We do not elaborate further on this degeneracy breaking since the addition of any of CMB lensing, BAO, or uncalibrated supernova measurements constrains neutrino masses sufficiently tightly that the alterations of early ISW effects are negligible.  

\subsection{CMB Lensing}

As CMB photons travel from the last-scattering surface, their paths are distorted by gravitational potential gradients perpendicular to their momenta. Gradients in the resulting deflections lead to shearing, magnification, and demagnification of the otherwise statistically isotropic and Gaussian patterns of anisotropy. One can use the resulting departures from statistical isotropy to infer convergence maps and therefore the CMB lensing deflection power spectrum \citep{Zaldarriaga:1998ar, Hirata:2002jy, Okamoto:2003zw, Millea:2020cpw}. CMB lensing information is also contained in CMB power spectra inferred from the observed lensed maps. Since they are derived from both magnified regions in which angular spectral patterns are shifted to lower $\ell$, and demagnified regions with shifts to higher $\ell$, the acoustic features in the map are smeared, and the drop of power toward high $\ell$ caused by Silk damping is softened \citep{Lewis:2006fu}. 

That CMB measurements alone, including CMB lensing, can allow for tight constraints on $\mnu$, with errors comparable to the minimum value of 0.06 eV, was first pointed out by \citet{Kaplinghat:2003bh}. The impact of neutrino mass on CMB lensing is best explained with reference to Fig.~\ref{fig:Hubble_response}. Increasing $\mnu$ leads to an increased expansion rate at $z \gtrsim 1$ which slows down growth on scales below the neutrino free-streaming length, as a faster expansion rate inhibits the transport of matter from underdense regions to overdense regions. Increasing mass also decreases the neutrino free-streaming length; on scales above the free-streaming length the suppressive effect of the faster expansion is mitigated by the clustering of the neutrinos themselves, as they source large-scale gravitational potentials. The suppression of growth at $z \gtrsim 1$ leads to a suppression of the matter power spectrum today and of the CMB lensing power spectrum, on scales below the neutrino free-streaming length. 

\begin{figure}[t]
    \centering
    \includegraphics{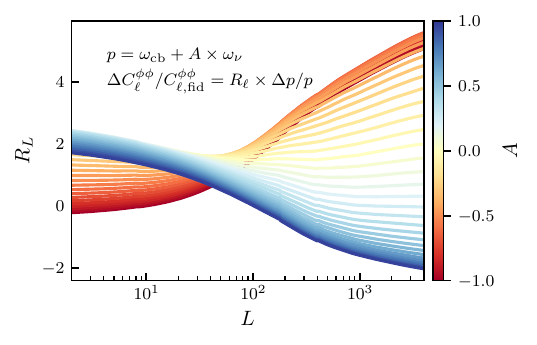}
    \caption{Lensing power spectrum responses to changes in linear combinations $\omcb+A\omnu$, as a function of angular scale and at high CMB likelihood. We see here that the linear combination to which $C_L^{\phi \phi}$ is most sensitive at low $L$ is one with $A > 0$, and at higher $L$ it is negative.}
\label{fig:clpp_response} 
\end{figure}

The scale dependence of this sensitivity is shown in Fig.~\ref{fig:clpp_response}. This figure shows the response in lensing power $\Delta C_L^{\phi \phi} / C_L^{\phi \phi}$ to changes in different linear combinations $\omcb + A\times \omnu$, as a function of angular scale. To compute these response curves, we take samples from a PR3, $\lcdm + \mnu$ chain and fit the linear response factor $R_L$ at each $L$ defined via:

\begin{equation}
    \frac{\Delta C_L^{\phi \phi}}{C_{L,{\rm fid}}^{\phi \phi}} = R_L \times \frac{\Delta \left(\omcb + A\times\omnu\right)}{ \left(\omcb + A\times\omnu\right)_{\rm fid}}.
\end{equation}
We scan a range of values for $A$ and do a fit for each value. By fitting to parameter variations taken from chain samples, we ensure that the responses shown are the ones that stay at high primary CMB likelihood.

This figure shows us which linear combination produces the largest response at a given $L$. We see that the linear combination giving the largest response at small $L$ has $A\approx 0.2$. This value is positive because neutrinos cluster like cold dark matter at these scales, which are larger than the free-streaming length. At smaller scales (larger $L$), the maximal-response linear combination asymptotes at $A \approx -0.65$ — the negative sign reflecting the fact that $\omcb$ and $\omnu$ have opposite effects on the lensing power at these scales. The cross-over happens in the range $30 \lesssim L \lesssim 100$, which approximately coincides with the neutrino free-streaming scale at redshifts of $z \sim 2-10$, where the CMB lensing kernel has support.

When only considering CMB power spectra, the lensing-induced peak smearing is the dominant effect contributing to neutrino mass bounds, and makes CMB data sensitive to neutrinos less massive than $m_{\nu} \simeq 0.6 \text{ eV}$ \citep{Bertolez-Martinez:2024wez}. Lensing allows one to break the background-level degeneracy with $\omega_{\Lambda}$ and place constraints on $\mnu$ with only CMB data. However, it also means that inferences of $\mnu$ are sensitive to anything that impacts the lensing inference, in particular the well-known $A_{\rm 2pt}$ (or, $A_L$) anomaly in Planck PR3 data \citep{Planck:2018vyg}. Indeed, as noted by many authors \citep[e.g.][]{Allali:2024aiv, Naredo-Tuero:2024sgf, Green:2024xbb, DiValentino:2021imh}, the $A_{\rm 2pt}$ anomaly contributes to the tight bounds on $\mnu$ from Planck data, because the reduction in lensing stemming from positive $\mnu$ results in a ``peak sharpening" that is out of phase with the oscillatory residuals in PR3 data, within \lcdm, that resemble peak smearing.

Lensing information also impacts the inference of $\omcb$, which is particularly relevant when combining with BAO data. Lensing power increases with increasing $\omcb$, mainly due to the higher matter-to-radiation density ratio when a perturbation mode of a given wavelength crosses the horizon. The greater this ratio, the less gravitational potential decay occurs. In the \lcdm\ model, the final constraint on $\omcb$ is a balance between fitting early universe effects (such as radiation driving and the associated boost in power), and this lensing-related effect impacting the CMB at later times. When $A_{\rm 2pt}$ is marginalized over, so that the lensing information in the CMB power spectra is effectively lost, the dark matter density inferred from the Planck PR3 data is lowered \citep{Planck:2018vyg}. We will return to this fact, and its relevance to our findings, in Section~\ref{sec:results}.

\subsection{BAO}

The sound horizon is also imprinted on the distribution of matter, and therefore the galaxy distribution. Using data from galaxy surveys, it is possible to infer
\begin{equation}
\theta_s(z) = \rs / D_A(z) \text{ and } \Delta z_s(z) = \rs / H(z)
\end{equation}
which correspond to the sound horizon scale in the transverse and line-of-sight directions \citep[e.g.][]{BOSS:2016wmc, eBOSS:2020yzd, DESI:2024uvr, DESI:2024lzq}. Here, $\rs$, is the sound horizon at the end of the baryon drag epoch. In the \lcdm\ model, even with extension to free $\mnu$, constraints on these parameters at multiple redshifts can be compressed losslessly into just two constraints. One is typically taken to be $\Omega_{\rm m}$, which completely controls the (un-normalized) shape of both $H(z)$ and $D_A(z)$. Here, $\Omega_{\rm m}$ is the present fraction of the energy density in non-relativistic matter, including the neutrino rest-mass density. The other is typically a constraint on $\rs h$, where $H_0 = 100\ h$ km/sec/Mpc. This second parameter pins down the normalizations allowing one to relate the un-normalized $H(z)$ (specified by $\Omega_{\rm m})$ to both $\theta_s(z)$ and $\rs H(z)$.

Inference of the physical matter density $\omm$, instead of $\Omega_{\rm m}$, would be much more valuable in terms of determining the neutrino mass because it could be directly compared to the CMB-inferred $\omcb$, with any difference attributable to the neutrinos. For this reason, the reparameterization  from ($\Omega_{\rm m}$, $\rs h$) to ($\Omega_{\rm m}$, $\omm \rs^2 $) has a significant advantage for understanding the impact of BAO constraints on $\mnu$ -- particularly when combined with CMB constraints \citep{Loverde:2024nfi}. Given some fiducial $\omega_{\rm b}$ and $\omega_{\rm c}$, one can then calculate $r_{\rm s}$ and then transform $\omega_m \rs^2$ to 
\begin{equation}
\label{eqn:xaxistransformf}
    \mnu = \frac{\omm \rs^2 - \omega_{\rm cb,f} r_{\rm d,f}^2} {r_{\rm d,f}^2} \times 93.14 \ {\rm eV}
\end{equation}
since $\omega_{\rm m} = \omega_\nu + \omcb$ and $\mnu = \omnu \times$ 93.14 eV. This transformation is, in a sense, comparing the matter density today of constituents that were non-relativistic prior to recombination, as inferred from CMB data, to the matter density inferred from BAO observations, after CMB calibration of $\rs$. In Section~\ref{sec:background_analysis}, we will use a similar (but subtly different) approach.

By comparing the constraints from the CMB BAO angle $\thetas$ and the galaxy survey BAO constraints (as inferred from both DESI and SDSS) in this plane, \citet{Loverde:2024nfi} found that they overlap primarily at values of $\omm \rs^2$ that are {\em less} than the best-fit $\omcb \rs^2$ as inferred from Planck CMB data. (See their Fig.~3 and/or our Fig.~\ref{fig:mnu_Omegam}). This is, of course, the opposite of what one would expect given $\mnu > 0$.

\subsection{Complementary Roles of BAO and CMB Lensing}

\begin{figure}[t]
    \centering
    \includegraphics{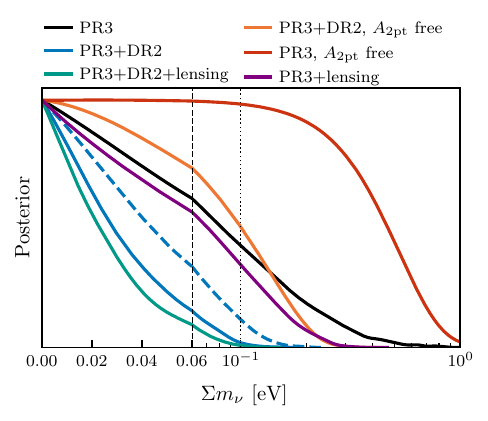}
    \caption{Constraints on $\mnu$ from different datasets, which are described in Table~\ref{tab:data_glossary}. The blue dashed line is the constraint one obtains from our signed neutrino mass model and background-only likelihoods  (both described in Section~\ref{sec:background_analysis}) with a prior $\mnue \geq 0$ eV.}
\label{fig:mnu_constraints_physical} 
\end{figure}

What is the relationship between BAO and CMB lensing as they pertain to measuring the neutrino mass? We now address this question, and beginning with a misconception. It is sometimes claimed that the main role of BAO data is in tightening the primary CMB-based predictions for the lensing amplitude, mainly through sharpening constraints on $\omega_{\rm c}$, so that any suppression (or enhancement) can be clearly detected \citep[e.g.][]{BOSS:2014hhw}. While this does indeed happen in the case that lensing and BAO data are used jointly, we see in Fig.~\ref{fig:mnu_constraints_physical} that adding BAO data to primary CMB data leads to significantly tighter $\mnu$ constraints regardless of how CMB lensing is treated. Clearly, BAO data cannot be purely a means to make better use of CMB lensing information.

This misconception may have arisen because the impact of neutrino mass on the density of non-relativistic matter is so small. It is indeed small, but as explained by \citet{Loverde:2024nfi} the interaction with $\thetas$ magnifies its importance, as can also be seen in Fig.~1 of \citet{Pan:2015bgi}. If one varies the expansion rate in a manner that keeps $\thetas$ fixed (i.e., at high CMB likelihood), the tiny contribution to the expansion rate from $\mnu$, because it persists over so much comoving distance on our past light cone, gets amplified: the cosmological constant has to make up for it over a much smaller co-moving distance. The low-redshift feature to which the BAO data are sensitive is best fit in \lcdm\ (at the $\thetas$ value preferred by Planck) via a high value of the cosmological constant, which is equivalent to a much more moderate decrease to the value of the matter density. It is this increase to the cosmological constant (with $\delta \omega_\Lambda \simeq - 13 \delta \omnu$ for masses near the minimum) that provides for the unusually high degree of sensitivity of the BAO data to small shifts in $\omega_{\rm m}$. 

Because CMB lensing and BAO data are sensitive to different linear combinations of $\omcb$ and $\omnu$, when combined with primary CMB data, they complement each other as probes of the neutrino mass. BAO data, once the sound horizon is calibrated from the CMB and with a precision boost from $\thetas$, place a constraint on the sum $\omm = \omcb + \omnu$. On the other hand, the linear combination of $\omcb$ and $\omnu$ that is best constrained by CMB lensing depends on the angular scales being probed. When probing scales smaller than the neutrino free-streaming scale, the best constrained linear combination $\omcb + A\omnu$ has $A<0$. The difference between the linear combinations constrained by BAO and CMB lensing data helps to isolate the value of $\omnu$.

\begin{figure}[t]
    \centering
    \includegraphics{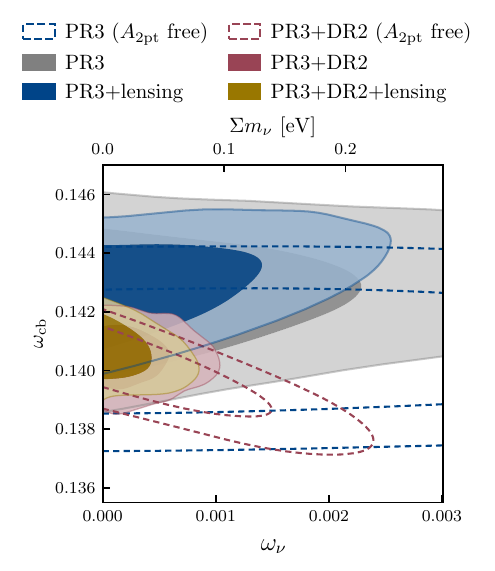}
    \caption{Constraints in the $\omnu-\omcb$ plane reveal the complementary nature of CMB lensing and BAO measurements for determining $\omega_\nu$ or, equivalently, $\mnu$. The datasets used here are described in Table~\ref{tab:data_glossary}. Dashed, unfilled contours do not contain any lensing information. Primary CMB data, with lensing information removed, do not constrain $\mnu$ for the mass range shown. When combined with primary CMB data, BAO data are sensitive to the sum $\omnu+\omcb$ while lensing data are sensitive to a combination that is close to the difference $\omcb-0.7\omnu$.}
\label{fig:omnu_omcb} 
\end{figure}

This interaction between datasets can be illuminated by considering the $\omega_{\rm cb} - \omega_\nu$ plane. We show this plane, along with constraints from different data combinations, in Fig.~\ref{fig:omnu_omcb}. The orientations of contours in this plane (by which we mean the slopes of their major axes) tells us which linear combination of $\omcb$ and $\omnu$ is best constrained by a given data combination. Primary CMB data, once the lensing information has been removed by marginalizing over $A_{\rm 2pt}$, is not sensitive to $\omnu$ for the range shown. They therefore constrain $\omcb - 0\times\omnu$, and the posterior contours shown in dashed blue are a horizontal band.

If BAO data are then included, the resulting posterior contours are approximately oriented along a line of constant $\omcb + \omnu$, as just discussed. We show this constraint in dashed red. With this data combination, we can see that the primary CMB (with lensing information removed) contributes little to the combined constraint on $\omcb$, which comes almost entirely from the BAO data. However, the primary CMB constraint on $\omcb$ cuts off the BAO degeneracy along $\omcb + \omnu$, preventing the dashed red contour from continuing to higher $\omnu$ and lower $\omcb$ values.

To understand the orientation of the contours involving lensing effects, we can make reference to Fig.~\ref{fig:clpp_response}. With the gray contour, we show constraints using primary CMB data including two-point lensing effects. The peak smearing in CMB power-spectra is predominantly sourced from scales near the peak of $C_L^{\phi \phi}$. Taking $L=50$ as a representative scale for this, we find that $A\approx-0.18$, which agrees well with the orientation of the gray contour along a line of constant $\omcb - 0.17 \omnu$. Lensing reconstructions probe $C_L^{\phi \phi}$ to smaller scales, so we expect greater sensitivity to $\omnu$. Taking $L=350$ as a representative scale probed by lensing reconstructions, we find $A\approx-.6$, which likewise agrees well with the orientation of the blue contour along a line of constant $\omcb - 0.7\omnu$. Because these lensing directions are anti-aligned with the CMB+BAO constraint on $\omcb+\omnu$, using the two probes leads to tight constraints on $\omnu$, as shown in the gold contour in Fig.~\ref{fig:omnu_omcb}.

Fig.~\ref{fig:omnu_omcb} also illustrates the impact of the higher $\omcb$ value preferred by lensing data. Along with changing which linear combination of $\omcb$ and $\omnu$ is constrained, including lensing also shifts constraints towards higher values of $\omcb$ for fixed values of $\omnu$. One implication of this is that, in the $\lcdm\ +\mnu$ model space, the lensing excess and matter-density deficit signatures are closely related. The low matter density preferred by BAO data when combined with primary CMB data reduces somewhat the expected lensing power, which contributes to the excess lensing problem. Similarly, the lensing reconstruction (and peak-smearing in the CMB power spectra) both prefer higher $\omega_{\rm c}$, and therefore $\omega_{\rm cb}$, which worsens the matter-density deficit.

We find it interesting that we can use Planck primary CMB data to set expectations for each of these probes, for a continuum of values of $\mnu$, and in both cases what is observed is the opposite of expectations assuming $\mnu> 0$.

\section{Background preference for $\omm < \omcb$}\label{sec:background_analysis}

In this section we quantify the preference, noticed by \citet{Loverde:2024nfi}, of BAO plus CMB data for a total non-relativistic matter density today that is less than the sum of baryonic and cold dark matter density today. We do so by defining a toy model with a signed neutrino mass parameter that can take on negative values, to allow for the possibility that $\omm < \omcb$. Such a difference could then be interpreted as a ``negative neutrino density" $\omnue \equiv \omm - \omcb$. This is similar in spirit to other signed neutrino mass implementations \citep{Craig:2024tky, Green:2024xbb, Elbers:2024sha, DESI:2025ejh}, but one which only models the impact of the neutrino on the background evolution. We first discuss our implementation of the model, as well as our method for comparing to data while only modeling the background evolution. We then present the results of this analysis, which show a preference from CMB+BAO data for a matter-density deficit. 

\subsection{Model}\label{sec:model}

Here we construct a phenomenological model of the background evolution with a signed neutrino mass parameter that is allowed to be negative. We assume two species of massless neutrino, which behave in the standard way, and one neutrino with this signed mass.

As with other species, neutrinos enter into the Friedmann equation governing the background evolution:

\begin{equation}
    h^2(z) = \omcb(z) + \omnu(z) + \omega_\gamma(z) + \omega_{\rm de}(z),
\end{equation}
where $h(z) \equiv H(z)/100 {\rm\, km/s/Mpc}$, $\omega_\gamma$ is the energy density of photons (which is well determined by local measurements of the CMB temperature \citep{Fixsen:1996nj}), and $\omega_{\rm de}(z)$ is the dark energy density, which we take here to be a constant $\omega_{\rm de}(z) \equiv \omega_{\Lambda}$. The energy density in neutrinos $\omnu(z)$ contains contributions from both the two species of massless neutrino as well as the massive species.

To phenomenologically extend the impact of massive neutrinos to negative values, we define a new parameter $\mnue$. In terms of the signed mass parameter $\mnue$, the Friedmann equation is written:

\begin{equation}
    \begin{split}
        h^2(z; \mnue) &\equiv h^2(z; 0) + \mathrm{sgn}(\mnue)\times \\ &\left(h^2(z; \mnu = |\mnue| ) - h^2(z;  0) \right)    
    \end{split}
\end{equation}
In other words, the effect of a negative signed mass on the expansion rate is identical to that of an equal-in-magnitude positive, physical mass but with opposite sign. When computing the expansion rate for massless neutrinos in the above expression (the $h^2(z;0)$ terms) we increase $N_{\rm eff}$ to $3.044$.

To illustrate this, in Fig.~\ref{fig:Hubble_response} we show the expansion history relative to the expansion history with massless neutrinos for selected values of $\mnue$, where we have also adjusted $\omega_{\Lambda}$ so as to keep $\thetas$ fixed. The impact of positive $\mnue$ is identical to that of physical neutrinos with $\mnu = \mnue$, with an increase in the expansion rate during matter domination at $z \gtrsim 1$ and a corresponding decrement during DE domination (per the discussion in Section~\ref{sec:theory}). For negative values of $\mnue$ the impact is reversed, with a decrement during matter domination, and an increase at $z \lesssim 1$.

We note that there are multiple ways to define a signed neutrino mass parameter at the background level. The model described here reverses all background-level effects of neutrino mass for negative values of the signed mass parameter. As such, the redshift dependence of $H(z; \mnue)$ is symmetric around $\mnue=0$, as can be seen in Fig.~\ref{fig:Hubble_response}. We have also tested a model with different redshift dependence, and found parameter shifts of around $0.1\, \sigma$.
\footnote{In this model, we  defined the ``matter-like" component of the neutrino energy density to be $\rho_{\nu, {\rm matter}} = \rho_\nu - 3p_\nu$. In the Friedmann equation, an effective neutrino energy density was included as $\rho_{\nu, {\rm effective}} \equiv {\rm sgn}(\mnue)\rho_{\nu, {\rm matter}} + 3p_{\nu}$. Here, $\rho_\nu$ and $p_\nu$ are respectively the energy density and pressure of massive neutrinos.}

In addition to the background evolution, we need to model the recombination process in order to compute quantities such as $\thetas$ or $D_A^\star$, which depend on $z_\star$. We use the HYREC-2 recombination code \citep{PhysRevD.83.043513, PhysRevD.102.083517} for these calculations. With the ionization fraction obtained from HYREC-2, we compute $z_\star$ as the redshift at which the optical depth for Thomson scattering crosses unity, defined by

\begin{equation}\label{eqn:optical_depth}
   \tau(z_\star) \equiv \int_0^{z_\star} \frac{X_{\rm e}(z) n_{\rm H}(z) \sigma_T}{H(z)(1+z)}dz = 1.
\end{equation}
Here, $X_{\rm e}(z)$ is the ionization fraction, $n_{\rm H}(z)$ is the number density of hydrogen nuclei, and $\sigma_T$ is the Thomson scattering cross section. The redshift of the end of the baryon drag epoch, $z_{\rm d}$, is defined similarly but with an additional factor of $R(z) = 3\omega_{\rm b}(z) / 4 \omega_{\gamma}(z)$ in the denominator. In combination with Equation~\ref{eqn:rs_da}, we can then compute $r_s^\star, r_s^{\rm d}, D_A^{\star},\,\text{and}\, \thetas$ as needed.

For calculations involving the expansion rate at $z<z_\star$ (e.g. $D_A(z)$, $H(z)$, $z_\star$), we use the signed neutrino mass model just described. For calculations involving the expansion rate at $z>z_\star$, we calculate $H(z)$ assuming massless neutrinos, and $N_{\rm eff} = 3.044$. The only exception to this rule is in the calculation of the recombination history, which bridges these regimes; for this, we likewise assume massless neutrinos.

The reason for these choices is twofold. First, this ensures we are only modeling the effect of neutrino masses we intend to model: their impact on the distance to last scattering. The second reason has to do with how we constrain this model, which we will explain more thoroughly in Section~\ref{sec:data}. In short, we use ``background-only" likelihoods that are created from the marginal posteriors $\mathcal{P}(\omega_{\rm b}, \omega_{\rm c}, \thetas)$ obtained from CMB analyses which assume fixed $\mnu = 0.06 \text{ eV}$. For values of $|\mnue|\gtrsim0.6\, {\rm eV}$, however, the impact of the signed neutrino mass parameter on the background expansion is large enough to affect the inference of these quantities from the primary CMB, and this would not be reflected by our background-only likelihoods. The above choices help to ensure that the model does not deviate too strongly from $\lcdm$ at early times, consistent with the assumptions relevant to the creation of the background-only likelihoods.\footnote{We stress that this second point is only potentially relevant for $|\mnue|\gtrsim0.6\, {\rm eV}$, and all of the constraints we quote below are much below this scale.}

\subsection{Data}\label{sec:data}

\begin{table*}[t]
\caption{\label{tab:data_glossary}
Glossary of data sets and background-only likelihoods labels used in this work. The background-only likelihoods are constructed according to the the procedure outlined in Section \ref{sec:data}, using the data listed in the table.}
\begin{ruledtabular}
\begin{tabular} {l |l}
\textbf{Name} & \textbf{Description}  \\
\hline
\makecell[tc]{PR3} & \makecell{Planck 2018 \texttt{Commander} low-$\ell$ TT, \texttt{SimAll} low-$\ell$ EE, and $\texttt{Plik}$ (nuisance parameter marginalized) \\ high-$\ell$ TTTEEE likelihoods \citep{Planck:2019nip}} \\
\makecell[tc]{PR4} & \makecell{Planck 2018 \texttt{Commander} low-$\ell$ TT, \texttt{LoLLiPoP} low-$\ell$ EE, and \texttt{HiLLiPoP} high-$\ell$ TTTEEE \citep{Tristram:2023haj} likelihoods \\ using the Planck PR4 NPIPE data \citep{Planck:2020olo}}\\
\makecell[tc]{P-ACT} & \makecell{Joint ACT DR6  CMB-only likelihood (\texttt{DR6-ACT-lite}) \citep{2025arXiv250314452L, 2025arXiv250314451N}, combined with Planck data in the manner \\ recommended by the ACT collaboration.} \\
\makecell[tc]{DR2} & DESI DR2 BAO measurements \citep{DESI:2025zgx}.\\
\makecell[tc]{lensing} & The unified ACT DR6, SPT-3G, and Planck PR4 lensing likelihood \citep{SPT-3G:2025zuh,SPT-3G:2024atg,ACT:2023dou,carron_cmb_2022} \\
\hline
\makecell[tc]{PR3-b} & \makecell{Background-only likelihood created using PR3 \texttt{plikHM\_TTTEE\_lowl\_lowE} chains. \footnote{\url{https://pla.esac.esa.int/}}}\\
\makecell[tc]{NoLens-b} & \makecell{Background-only likelihood created using using PR3 \texttt{Alens\_plikHM\_TTTEE\_lowl\_lowE} chains \footnote{\href{https://pla.esac.esa.int/}{Ibid.} } } \\
\makecell[tc]{PR4-b} & Background-only likelihood created using PR4 \texttt{hlpTTTEEE\_lowT\_lolE} chains. \footnote{\url{https://portal.nersc.gov/cfs/cmb/planck2020/}}\\
\makecell[tc]{P-ACT-b} & Background-only likelihood created using P-ACT \texttt{p-actbase\_lcdm\_camb} chains. \footnote{\url{https://portal.nersc.gov/project/act/dr6.02/chains/}}\\
\makecell[tc]{DR2-cb} & \makecell{The DR2 likelihood indicated above, in combination with a joint prior on $\omega_{\rm c}$ and $\omega_{\rm b}$ from one of \\ the CMB likelihoods (the specific one depending on context).}
\end{tabular}
\end{ruledtabular}
\end{table*}

\begin{figure}
    \centering
    \includegraphics{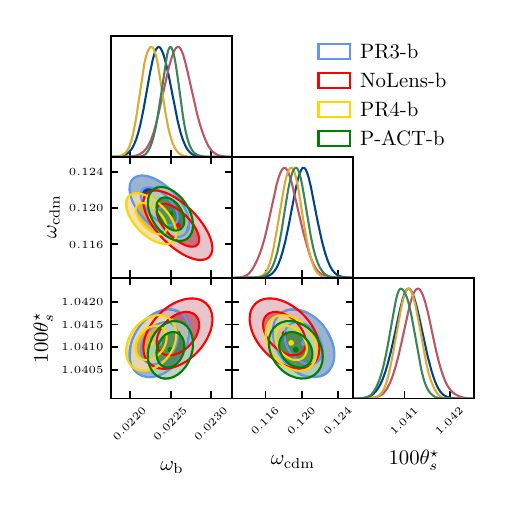}
    \caption{One and two-dimensional marginal posterior distributions for the 'background' CMB-derived parameters (see text), for the 'Background-only' (B) likelihoods. 
    The B likelihoods are Gaussian fits to the full likelihoods, whose two-dimensional marginal posterior distributions are shown here as filled contours. The Gaussian fits are sufficiently good approximations that no differences can be seen here. The likelihoods used are described in Table \ref{tab:data_glossary}. All contour plots are made using the GetDist package \cite{Lewis:2019xzd}.}
\label{fig:likelihoods} 
\end{figure}

\begin{figure*}[t]
    \centering
    \includegraphics{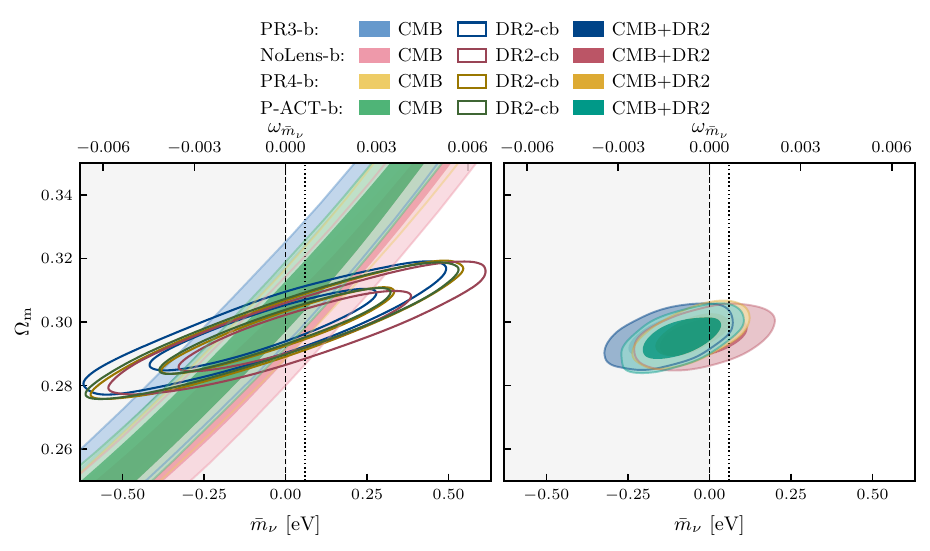}
    \caption{\textit{Left panel}: Constraints from CMB and BAO data, considered separately, using the three background-only likelihoods. For the CMB constraint, we use the full three parameter background-only likelihood, whereas for the BAO-cb constraints we use BAO likelihoods along with a joint Gaussian prior on $\omega_{\rm b}$ and $\omega_{\rm c}$ with covariance obtained from the background-only likelihood. The top axis shows the signed neutrino energy density using $\mnue / 93.14 \text{ eV} = \omnue$. \textit{Right panel}: The constraints using the background-only CMB likelihood jointly with BAO data.}
    \label{fig:mnu_Omegam}
\end{figure*}

Since we only model the background evolution with our signed neutrino mass model, we cannot use full CMB likelihoods for constraints. Instead, in this subsection, we use constraints obtained from CMB analyses with the full likelihood to create ``background-only" likelihoods which encapsulate the CMB information about the post-recombination background evolution. It is these likelihoods that we use in the subsequent analysis.

As discussed in Section \ref{sec:theory}, of the six \lcdm\ parameters, the physical baryon and dark matter densities, along with the angular scale of the sound horizon, $\vec{\theta} = \left(\omega_{\rm c}, \omega_{\rm b}, \theta_s^\star \right)^{\intercal}$, are well determined from the CMB and are sufficient (within \lcdm) to determine the post-recombination background evolution. We use the marginal posteriors of these parameters, coming from full CMB analyses, to construct three background-only likelihoods as follows. 

Given samples from the posterior distribution under some dataset and model, we construct a parameter covariance matrix $\Sigma \equiv \text{cov}\left(\vec{\theta},\vec{\theta} \right)$. We use this covariance matrix to form a Gaussian approximation to the marginal posterior in this 3-dimensional subspace. We choose the mean value of these parameters as the ``data vector", so that the background-only likelihood is

\begin{equation}
    \ln{\mathcal{L}} = \left(\vec{\theta} - \vec{\theta}_{\rm data} \right)^\intercal \Sigma^{-1}  \left(\vec{\theta} - \vec{\theta}_{\rm data} \right).
\end{equation}
To distinguish these likelihoods from the full likelihoods from which they are constructed, we denote them with a ``-b" suffix. The ``background-only" nomenclature is meant to convey that these likelihoods encapsulate the CMB constraint on the post-recombination background evolution. This approach to compressing CMB information has been used in the past, e.g. in \citep{BOSS:2014hhw, 2025arXiv250314452L}.

The data we use to create these likelihoods is summarized in Table~\ref{tab:data_glossary}. For our baseline CMB data, we use Planck PR3 data, and construct the baseline background-only likelihood from posterior samples assuming \lcdm. However, the lensing information in the CMB two point functions plays an important role in setting expectations for $\omcb$, which is relevant to our analysis. We therefore also create a likelihood from posterior samples using PR3 data, after marginalizing over $A_{\rm 2pt}$. We also create a likelihood from Planck PR4 data, where the $A_{\rm 2pt}$ excess is not as significant. Finally, we consider one likelihood created using a combination of Planck and ACT data, which is the most constraining CMB dataset we consider here. These background-only likelihoods are shown in Fig.~\ref{fig:likelihoods}.

We also define a set of ``BAO-cb" likelihoods, capable of constraining $\mnue$. Each one is an extension of the BAO likelihood combination listed in Table~\ref{tab:data_glossary} to include a joint posterior of $\omega_{\rm b}$ and $\omega_{\rm c}$ obtained by marginalizing the CMB background-only likelihood over $\thetas$. With this extension, we can plot BAO constraints (with this CMB prior) directly in the $\Omega_{\rm m} - \mnue$ plane after marginalizing over uncertainties in $\omega_{\rm b}$ and $\omega_{\rm c}$. This is similar to the $\Omega_m -\omm \rs^2$ plane used by \citet{Loverde:2024nfi}, but allows us to marginalize over the CMB information instead of assuming fiducial values. Note that we use Equation~\ref{eqn:xaxistransformf}, with $\mnu$ replaced by $\mnue$, and $r_{\rm d,f}$ replaced with $r_{\rm d}$ to plot in this plane.
Since $\omega_{\rm m} r_{\rm d}^2$ can be determined from BAO data, and the remaining quantities can be determined from $\omega_{\rm b}$ and $\omega_{\rm c}$, the BAO-bc likelihoods are indeed capable of constraining $\mnue$, as desired.

\subsection{Results}\label{sec:results}
\begin{table}[t]
\caption{\label{tab:priors}
Priors used for the background-only and DDM analyses presented in this work. In the DDM analysis, $\Gamma$ is treated as a derived parameter following the relationship discussed in Section \ref{sec:DDM}. In both analyses, we assume two massless and one massive species of neutrino.}
\begin{ruledtabular}
\begin{tabular} {l|l|l}
\textbf{Model} & \textbf{Parameter} & \textbf{Prior}  \\
\hline
\makecell[cc]{\textbf{Background only}} & $\omega_\Lambda$ & $\mathcal{U}[0.001, 0.99]$ \\
& $\omega_{\rm b}$ & $\mathcal{U}[0.005, 1]$  \\
& $\omega_{\rm c}$ & $\mathcal{U}[0.001, 0.99]$ \\
& $\mnue$  & $\mathcal{U}[-1, 1]$  \\
\hline
\makecell[cc]{\textbf{DDM}} & $100 \theta_s^\star$ & $\mathcal{U}[0.5, 5]$ \\
& $\omega_{\rm b}$ & $\mathcal{U}[0.005, 1]$  \\
& $\omega_{\rm c}$ & $\mathcal{U}[0.001, 0.99]$ \\
& $\ln\left(10^{10}A_s \right)$ & $\mathcal{U}[1.61, 3.91]$ \\
& $n_s$ & $\mathcal{U}[0.8, 1.2]$ \\
& $\tau_{\rm reio}$ & $\mathcal{U}[0.01, 0.8]$ \\
& $\Sigma m_{\nu}$ & $\mathcal{U}[0.0, 3.0]$ \\
& $\omddm$ & $\mathcal{U}[0.0, 0.01]$\\
& $A_{\rm 2pt}$ & $\mathcal{U}[0, 10]$
\end{tabular}
\end{ruledtabular}
\end{table}

\begin{figure}[t]
    \centering
    \includegraphics{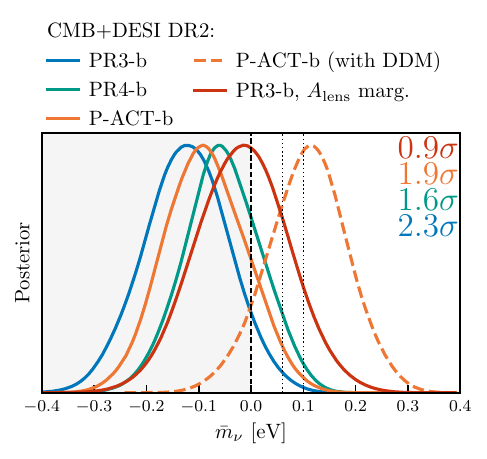}
    \caption{One-dimensional marginal posteriors for the signed neutrino mass parameter $\mnue$ given several different background-only CMB + BAO likelihoods. All cases assume \lcdm\ with extension to free $\mnue$. The curve labeled `fixed DDM' includes a contribution from decaying dark matter with fixed model parameters described in Section~\ref{sec:DDM}.}
    \label{fig:mnu_1d}
\end{figure}

We constrain the four-parameter signed neutrino mass model in an MCMC analysis with the \texttt{COBAYA} \citep{Torrado:2020dgo} sampler, using the reduced CMB and BAO likelihoods and BAO data listed in Table~\ref{tab:data_glossary}.\footnote{The code for our background model and likelihoods can be found at this link: \url{https://github.com/gplynch619/backgroundmnu}.} All of the following constraints include DR2 data. We consider chains as converged when the Gelman-Rubin statistic is at $R-1 < 0.01$. Parameter constraints are reported as 68\% confidence intervals.

We begin with a comparison between a result derived with physical neutrino masses and a result in our signed neutrino mass model, in the regime where they can be compared, namely for $\mnue\, > \, 0$. We run a chain with the PR3-b and BAO likelihoods, with a prior that $\mnue\, >\, 0$. The resulting constraint on $\mnue = \mnu$ is shown as the blue dashed line in Fig.~\ref{fig:mnu_constraints_physical}. The difference between the blue curves in this plot indicate that the compression to background-only likelihoods is not completely lossless. This is because there is lensing information in the primary CMB beyond that which is encapsulated by the value of $\omega_{\rm c}$ in the compressed likelihoods. We discuss this further in Appendix~\ref{app:lkl_comparison}.

We now move on to cases where $\mnue$ can be negative. With the baseline PR3-b likelihood, we find a preference for $\omm < \omcb$, with

\begin{equation}
    \mnue = \left(-0.123\pm 0.081\right) \text{ eV}.
\end{equation}
This excludes $0.06$ eV, the minimum mass in the NO hierarchy, at $2.3\, \sigma$. This likelihood includes the effects of lensing in the two point function, and correspondingly $\omega_{\rm c}$ adjusts towards higher values to fit the features in the Planck data driving the $A_{\rm 2pt}$ anomaly. To assess the impact of this, we can use the NoLens-b likelihood, where lensing information has been removed by marginalizing over $A_{\rm 2pt}$. In that case, we find a much weaker preference for a matter-density deficit, with

\begin{equation}
    \mnue = \left(-0.017\pm 0.089\right) \text{ eV}.
\end{equation}
This is completely consistent with $0.06\, \mathrm{eV}$, being only $ 0.87\, \sigma$ away. With the PR4-b likelihood, where $A_{\rm 2pt}$ is consistent with unity at the $0.7\, \sigma$ level, the constraint is
\begin{equation}
    \mnue = \left(-0.054\pm 0.074\right) \text{ eV}.
\end{equation}
which is $1.6\, \sigma$ away from the NO minimum mass.

Finally, using the P-ACT-b likelihood, we obtain a constraint of

\begin{equation}
    \mnue = \left(-0.088\pm 0.079\right) \text{ eV}.
\end{equation}
which is $1.9\, \sigma$ away from $0.06\, {\rm eV}$.
These posteriors are shown in Fig.~\ref{fig:mnu_1d}.

\subsubsection{Discussion}

In Fig.~\ref{fig:mnu_Omegam}, we demonstrate how the separate CMB-b and BAO constraints interact to produce this preference for a negative signed neutrino mass. In the left panel, we show the CMB-b and DR2-cb constraints separately in the $\mnue-\Omega_{\rm m}$ plane. As anticipated from the discussion in Section \ref{sec:theory}, when only CMB data are used there is a broad degeneracy in this plane that stems from the degeneracy between $\omnue$ and $\omega_\Lambda$ that keeps the distance to last scattering fixed. BAO data, when the sound horizon $\rs$ is calibrated from the CMB prior on $\omega_{\rm b}$ and $\omega_{\rm c}$, can directly constrain low-redshift distances. Combined with the value of $\omcb$ obtained from the CMB prior, this results in the unfilled contours in the left-hand panel. 

Examining these constraints separately helps to isolate the impact of the shifts in $\omega_{\rm cb}$ when marginalizing over $A_{\rm 2pt}$. The excess lensing-like smoothing that is present in PR3 data is not well fit within \lcdm: the fit improves by $\Delta \chi_{\rm eff}^2 \approx -10$ when $A_{\rm 2pt}$ is freed \citep{Planck:2018vyg}.\footnote{This improvement is not only due to fitting the lensing-like feature at $600 \lesssim \ell \lesssim 1500$, but also due to improved fits at large angular scales made possible by degeneracies between $A_{\rm 2pt}$ and other parameters \citep{Planck:2016tof, Planck:2018vyg}.} These improved fits have a lower matter density than is preferred in \lcdm, which can be seen in the reduced likelihoods shown in Fig.~\ref{fig:likelihoods}. We see from the right panel of Fig.~\ref{fig:mnu_Omegam}, and from the decreased tension with the NO minimum mass when $A_{\rm 2pt}$ is freed, that this shift in the inferred $\omega_{\rm c}$ impacts the significance of the $\omm < \omcb$ preference. One can also see how adding the $\thetas$ constraint significantly tightens the BAO constraint on $\bar m_\nu$.

In general, CMB datasets that prefer lower $\omcb$ lead to higher values of $\mnue$. This can be seen in both panels of Fig.~\ref{fig:mnu_Omegam}, where, for fixed values of $\Omega_{\rm m}$, the datasets with lower $\omcb$ (such as NoLens-b) are shifted to the right, towards higher $\mnue$. This can be understood by considering the constraint on $\omnue \equiv \omm - \omcb$. When the lensing amplitude is free, and the CMB prefers lower values of $\omcb$, the difference $\omm-\omcb$ required for consistency between CMB and DR2 data is lessened, resulting in a milder preference for $\omm < \omcb$. With the PR4 likelihoods that we use, the excess peak smearing in the power spectra is reduced, with $A_{\rm 2pt} = 1.039 \pm 0.052$ ($0.7\, \sigma$ from unity). Here, the inferred $\omcb$ is higher than it is when using PR3 data and marginalizing over $A_{\rm 2pt}$, as the dark matter density must still adjust (towards higher values) to fit the peak smoothing. The results using the P-ACT-b likelihood follow this general pattern as well; this data combination (which, when it is freed, finds $A_{\rm 2pt} = 1.080 \pm 0.042$) has a preference for $\mnue$ between PR3 and PR4, but with lower uncertainties.

The correlation between $A_{\rm 2pt}$ and $\omcb$, which arises due to the role $\omcb$ plays in the lensing of primary CMB power spectra, leads to one interesting conclusion: extensions to \lcdm\ that predict a higher lensing power (\lcdm+$A_{2\rm pt}$ being an example, albeit not a physical one) could potentially solve both the lensing excess and the matter-density deficit.

Finally, we compare our results with those of \citet{DESI:2025ejh}, who have also used a model with signed neutrino masses to assess tensions between CMB and DR2 data in the $\lcdm +\mnu$ model space. Their model reverses the impact of neutrino mass on all observables \citep{Elbers:2024sha} (instead of only the background expansion, as in this work), and they use the PR4 \texttt{CamSpec} high-$\ell$ likelihood of \citet{Efstathiou:2019mdh}. For this different model and CMB data combination, they find a $2.7\, \sigma$ preference for negative signed neutrino mass when combining with DR2. 

Using the PR4 \texttt{CamSpec} likelihood and freeing the amplitude of lensing, one finds a value of $A_{\rm 2pt} = 1.095 \pm0.056$ \citep{Rosenberg:2022sdy}, higher than the PR4 likelihood that we use. Therefore, we would expect that switching to this likelihood would increase the significance of the $\mnue < 0$ preference. Additionally, the \citet{DESI:2025ejh} model can respond to the information in the full likelihoods, including that which is lost during compression to the background-only likelihoods (see Appendix~\ref{app:lkl_comparison}). These factors contribute to the difference in significance between our results and those of \citet{DESI:2025ejh}.

\section{Case study: Decaying dark matter}\label{sec:DDM}

\begin{figure}[t]
    \centering
    \includegraphics{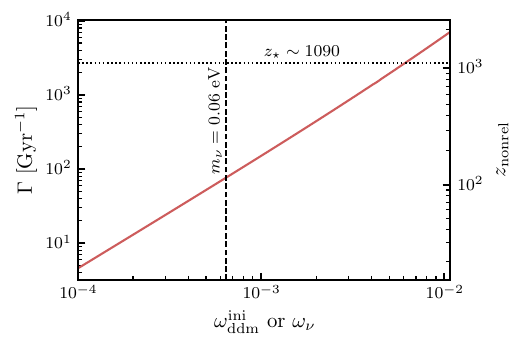}
    \caption{The relationship between decay rate $\Gamma$ and the initial decaying dark matter density parameter $\omega_{\rm ddm}^{\rm ini}$ in our reduced-dimensionality decaying dark matter parameter space (left-hand $y$ axis), and the relationship between a given neutrino energy density today $\omega_\nu$  and the redshift of the neutrino's relativistic to non-relativistic transition (right-hand $y$ axis). In our DDM model, the decay rate is a derived parameter, chosen such that a given $\omega_{\rm ddm}^{\rm ini}$ decays at the non-relativistic transition redshift for $\omega_\nu = \omega_{\rm ddm}^{\rm ini}$ assuming one species of massive neutrino. This ensures the DDM is degenerate with neutrino mass in its effect on the background expansion. For reference, the density corresponding to $m_{\nu} = 0.06$ eV is marked, as is the redshift (and corresponding decay rate) of last scattering.}
    \label{fig:omega_gamma_relation}
\end{figure}

\begin{figure}[t]
    \centering
    \includegraphics{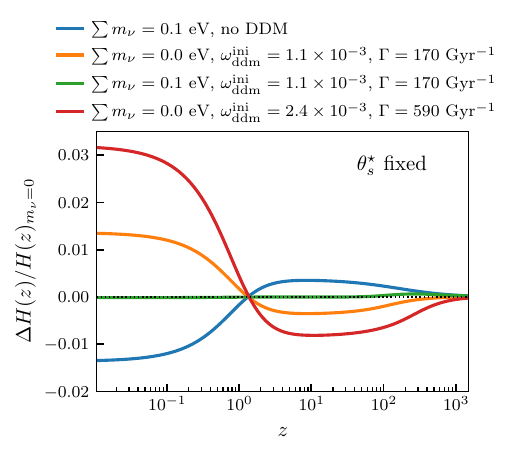}
    \caption{Changes to the expansion rate caused by neutrino mass can be mimicked by decaying dark matter. Here we show several departures from the case of zero neutrino mass and zero decaying dark matter. The decaying dark matter parameter values for the green curve can be used to compensate for $\Sigma m_\nu = 0.1$ eV.}
    \label{fig:Hz_ddm}
\end{figure}

In the previous section, we established that at the background level, the combination of CMB and BAO data shows a preference for a deficit in the matter density, $\omm < \omcb$, once freedom has been introduced to allow for that possibility. The phenomenological model we used served to quantify the problem, but is, of course, not a solution.   

We now consider a model of decaying dark matter (DDM) with dark radiation (DR) decay products, which, similar to the signed neutrino mass model of the previous section, can decrease the comoving matter density after recombination. We show that this model is indeed similar to the signed neutrino mass model at the background level. We then place constraints in a model space where the neutrino mass is free to vary, as is the initial amount of decaying dark matter. We consider this model a case study illustrating how the $\omm {\, }< {\, }\omcb$ preference could be explained, without solving the excess lensing problem. Alternatively, this case study illustrates the difficulties in explaining both signals with a physical model.

Decaying dark matter (DDM) models have been studied as potential solutions to both the $S_8$ and Hubble tensions \citep[e.g.,][]{Enqvist:2015ara, Davari:2022uwd, PhysRevD.105.103512, Haridasu:2020xaa, Pandey:2019plg, Chudaykin:2017ptd, McCarthy:2022gok}. CMB, BAO, and other large-scale structure data have been used to constrain the amount and lifetime of DDM  \citep[e.g.][]{Ichiki:2004vi, Audren:2014bca, Poulin:2016nat, Xiao:2019ccl, Nygaard:2020sow, Holm:2023uwa}, and a general conclusion from these studies is that DDM models cannot resolve the aforementioned tensions without additional modifications to the cosmological model. These conclusions are also supported by frequentist analyses, which do not depend on the choice of priors, and are not susceptible to volume effects \citep{McCarthy:2022gok, Bringmann:2018jpr, Holm:2023uwa}.

For our case study, we use the DDM model currently implemented in $\texttt{CLASS}$ \citep{2011arXiv1104.2932L, 2011JCAP...07..034B}, which allows one to specify the initial density of DDM, $\omddm$, as well as the decay rate $\Gamma$. The evolution of the DDM density, and the density of its DR decay product, is given by \citep{Kang:1993xz, Audren:2014bca, Poulin:2016nat}
\begin{align}
    \dot \rho_{\rm ddm} &= -3 \frac{\dot a}{a} \rho_{\rm ddm} - \Gamma \rho_{\rm ddm} \, {\rm and} \label{eqn:ddm_1}\\
    \dot \rho_{\rm dr} &= -4 \frac{\dot a}{a}\rho_{\rm dr} + \Gamma \rho_{\rm ddm} \label{eqn:ddm_2}
\end{align}
where the $\dot \ $ indicates differentiation with respect to proper time. 
Decays become significant (and complete), on a timescale set by $t_{\rm ddm} \sim 1/\Gamma$. The parameter $\omddm$ is used to set the initial conditions, and is the initial comoving density of decaying dark matter. This model is a special case of the one introduced in \citep{Bringmann:2018jpr} and further studied in \citep{McCarthy:2022gok}.

Despite the attention they have received in relation to the $H_0$ and $S_8$ tensions, decaying dark matter models are less studied in the context of free $\mnu$. It was noted in \citet{BOSS:2014hhw} that DDM has a similar, but opposite, effect as massive neutrinos on the background evolution. \citet{Poulin:2016nat} explored this degeneracy further, and showed that it is nearly complete in its impact on the background expansion when two conditions are met. The first is that $\omddm = \omnu$, so that the contribution to $\omm$ from non-relativistic neutrinos is canceled by the decaying dark matter converting to radiation. The second condition is that the DDM decays at close to the same time as the neutrinos become non-relativistic, i.e. $\Gamma \approx 1/t(z_{\rm nr})$, where $z_{\rm nr}$ is given in Equation~\ref{eqn:znr}. Figure~\ref{fig:omega_gamma_relation} shows this relationship, where we have used the PR3 mean parameters to compute the redshift-time relationship, $z(t)$. For a given $(\mnu, z_{\rm nr})$, the corresponding pair $(\omddm, \Gamma)$ under this relationship will nearly completely cancel the effects of the neutrino mass in the background evolution. We show this explicitly in Fig.~\ref{fig:Hz_ddm}.

\citet{Poulin:2016nat} also demonstrated that this degeneracy is broken in CMB power spectra due to the impact that DDM has on gravitational lensing and large scale polarization, even when moving along the approximate background-preserving degeneracy. The polarization effects manifest as features on large scales; since they are not as important as the lensing effects we are about to discuss, we do not consider them further here. Lensing is affected in DDM models because of the conversion of non-relativistic, non-pressure supported matter to massless particles which do not cluster on sub-horizon scales. The result is that gravitational potentials decay, and lensing is suppressed, despite the reduction in expansion rate that, were it achieved without loss of clustering matter, would result in a boost to lensing.

It is not {\it a priori} clear to us if these degeneracy breaking effects are significant enough to exclude DDM models as solutions to the matter-density deficit. In this section, we investigate this possibility. Throughout, we enforce the approximate background-preserving relationship between $\omddm$ and $\Gamma$. We do so for two reasons. First, the DDM model space is known to suffer from volume effects, which we wish to avoid for the purpose of this case study. Treating $\Gamma$ as a derived parameter will reduce the dimensionality of our parameter space, which helps to mitigate these effects. Second, by enforcing this relationship we are considering an ``optimistic" scenario in which the background evolution is nearly unchanged from the $\mnu=0$, \lcdm\ case: if the model is nonetheless disfavored, we find it unlikely that it would become more favored by introducing changes at the background level.

\subsection{Background analysis}

Before performing an analysis in the DDM + $\mnu$ parameter space we just described, we first check if adding a fixed amount of DDM into the background analysis of Section \ref{sec:results} can completely remove the preference for $\omm < \omcb$. 
Specifically, we add fixed $\omega_{\rm ddm}(z)$ and $\omega_{\rm dr}(z)$ components to the Friedmann equations with values that follow from $\omddm = 2.4 \times 10^{-3}$ and $\Gamma = 590 \text{ Gyr}^{-1}$.

We show the resulting $\Delta H(z)/H(z)_{m_{\nu}=0}$ as the red curve in Fig.~\ref{fig:omega_gamma_relation}. We chose these values to result in an expansion history slightly more extreme than that preferred by P-ACT-b data, which in our signed neutrino mass model corresponds to $\mnue = -0.09 \text{ eV}$. Therefore, by construction, we expect a preference for positive neutrino mass, so as to recover the preferred expansion history (i.e. the green curve in Fig.~\ref{fig:Hubble_response}).

We compute $\omega_{\rm ddm}(z)$ and $\omega_{\rm dr}(z)$ by solving Equations~\ref{eqn:ddm_1} and \ref{eqn:ddm_2} given these initial conditions, which we do not sample over. The CMB constrains the dark matter density prior to recombination, so when including DDM in the analysis we treat the $\omega_{\rm c}$ constraint in our background-only likelihoods as a constraint on $\omega_{\rm dm}^\star$, the comoving dark matter density (including both stable and decaying particles) at last scattering.

Repeating the analysis using P-ACT-b, we find
\begin{equation}
    \mnue = \left(0.11\pm 0.07\right) \text{ eV}
\end{equation}
This posterior is shown in Fig.~\ref{fig:mnu_1d} as the dashed line. This confirms that, at least at the background level, the matter-density deficit could be explained by a decaying dark matter component along with massive neutrinos.

\subsection{Constraints on DDM + $\mnu$}

\begin{figure}[t]
    \centering
    \includegraphics{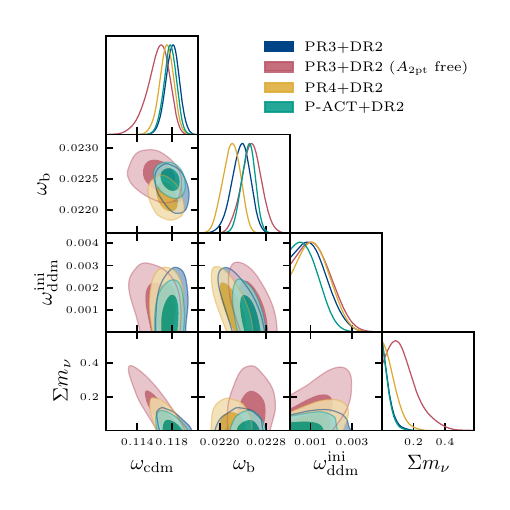}
    \caption{Constraints on density-related parameters in the \lcdm\ + $\Sigma m_\nu$ + decaying dark matter parameter space. These constraints are from the full CMB TT/TE/EE Planck likelihoods as described in Tab.~\ref{tab:data_glossary}. Note that lensing information allows for tight constraints on both the decaying dark matter and neutrino mass despite the geometric near-degeneracy shown in Fig.~\ref{fig:Hz_ddm}.}
    \label{fig:ddm_results}
\end{figure}

We now perform a full analysis of the DDM + $\mnu$ model using the CMB and BAO data listed in Table \ref{tab:data_glossary}. We sample over eight parameters, $\vec{\theta} = \{\omega_{\rm c}, \omega_{\rm b}, 100\theta_s^\star, \tau_{\rm reio}, n_s, \ln{\left(10^{10} A_s \right)}, \omddm, \mnu \}$, with $\Gamma$  treated as a derived parameter. Our priors are listed in Table \ref{tab:priors}. For CMB data, we separately consider both PR3, PR4, and P-ACT data, combining each with DR2. For PR3 data, we also consider the case where $A_{\rm 2pt}$ is free, to assess the impact that the lensing-induced peak smearing has on the constraints. We run each chain until the Gelman-Rubin statistic is at $R -1 < 0.04$. 

The constraints on the various physical matter densities are shown in Fig.~\ref{fig:ddm_results}. We see that even with the possibility of some amount of DDM, as long as $A_{\rm 2pt}=1$, the PR3+DR2 data still tightly constrain $\mnu$. When $A_{\rm 2pt}$ is allowed to freely vary, however, we see a shift in posteriors toward higher values of $\omddm$ and $\mnu$. This indicates that it is the lensing information present in the CMB data that leads to the still-tight constraints on $\mnu$: any amount of DDM, or any non-zero $\mnu$, will suppress lensing power, whereas the PR3 (in this case) data show a preference for enhanced lensing power. This same conclusion holds for the more constraining P-ACT data, wherein we also find tight constraints on the neutrino mass. This is consistent with the conclusions of \citet{McCarthy:2022gok}, who showed that for general DDM models, the fit to CMB power spectra predominantly degrades at intermediate multipoles where lensing effects are important. Freeing $A_{\rm 2pt}$ removes this constraint, and opens the expected degeneracy between $\omddm$ and $\mnu$. In the PR4 likelihoods that we use, the amount of peak smearing is consistent with \lcdm\ \citep{Tristram:2023haj}, so we might expect the $\omddm$-$\mnu$ degeneracy to more clearly manifest itself. Indeed, we see that switching to the PR4 data leads to a preference for non-zero DDM, and a posterior for $\mnu$ that supports more positive values (although still without a peak).

\section{\label{sec:discussion} Brief Discussion of Potential Solutions}

As we have seen with DDM, there are model extensions that can solve the matter-density deficit but do not solve the excess lensing problem, and vice versa. We now turn to a brief discussion of some of these possible solutions.

\subsubsection{Decaying dark matter}

We presented one extension, the DDM model, aimed at solving these problems, which we used as a case study. In this extension a fraction of the CDM is unstable and decays to dark radiation. At the background level, increasing DDM has nearly exactly the opposite impact as increasing $\mnu$. As a result, we found that with this model one could eliminate the matter-density deficit relative to minimal NO expectations. In this model, for any value of $\mnu$, the expansion rate is lower during the matter-dominated era than is the case for the corresponding \lcdm\ + $\mnu$ model.  This reduced expansion rate (which on its own would boost clustering) occurs due to conversion of clustering matter to radiation, so the lensing power is actually reduced, somewhat exacerbating the excess CMB lensing discrepancy rather than reducing it. 

\subsubsection{Primordial trispectrum}

Conversely, a potential solution proposed in \citet{Craig:2024tky} for the lensing power excess would probably not eliminate the matter-density deficit without significant additional model-building requirements. This solution involves the generation of primordial perturbations with a trispectrum that mimics the effects on the CMB from gravitational lensing. In this scenario CMB lensing reconstructions, which assume primordial Gaussianity and are dependent on the trispectrum, would be biased. With the right pattern of non-Gaussianity, that bias would show up as the observed excess of reconstructed CMB lensing power. 

Whether such a scenario could also address the matter-density deficit problem remains to be seen. If the new physics that alters the primordial four-point function also alters the primordial two-point function, then it could in principle change the inference of $\omega_{\rm cb}$ from CMB data in a way that resolves the matter-density deficit. However, if there is no appreciable change to the primordial two-point functions, and CMB lensing power is unchanged, then the inference of $\omcb$ from the primary CMB data will remain unchanged and the matter-density deficit discrepancy will remain. 

\subsubsection{Boosted lensing power}

We also mentioned previously that extensions to \lcdm\ that predict a higher lensing power, such as the phenomenological \lcdm+$A_{2\rm pt}$ model, could potentially solve both the excess lensing problem and the matter-density deficit problem. How such a model addresses excess lensing power is obvious. The potential for resolving the matter-density deficit discrepancy follows from the dependence of the inference of $\omega_{\rm cb}$ from primary CMB data. With a boosted lensing power spectrum, the $\omega_{\rm cb}$ inference drops,  reducing the significance of the matter-density deficit, although not outright restoring the preference for $\omm\, >\, \omcb$.

Finding a physical model that can boost the lensing power above the \lcdm\ result is challenging because in the \lcdm\ model the universe is dominated, for a large majority of the post-recombination doublings of the scale factor, by cold dark matter, which is already at the limit of zero pressure support and (practically) zero free-streaming length. Positive mean curvature ($\Omega_{\rm K} < 0$) to reduce the expansion rate would help, but the amount of tolerable curvature is tightly constrained by the combination of $\thetas$ and other geometrical probes sensitive to the low-redshift universe. \citet{Craig:2024tky} discuss the possibility of boosting growth via an attractive long-range dark matter self-interaction. As they point out, any implementation of this idea in a model will need to evade equivalence-principle limits. Additionally, other studies have shown that long-range dark matter self-interactions have a negligible impact on the lensing power spectrum, due to cancellations with background-level effects \citep{Archidiacono:2022iuu, Bottaro:2024pcb}.

\subsubsection{Boosted optical depth}

Multiple authors have recently noted that a value for the optical depth due to reionization, $\tau_{\rm reio}$, of around $\tau_{\rm reio} \sim 0.09$ could eliminate the matter-density deficit and lensing excess, due to the positive correlation with $A_s$ and the corresponding boost to lensing predictions \citep{Sailer:2025lxj, Jhaveri:2025neg, Allali:2024aiv}. As we have seen, increased lensing predictions generally result in lower $\omega_{\rm m}$, addressing the matter-density deficit. Interestingly, this is around the value that is preferred by current cosmological data, if the large scale polarization measurements from Planck are removed \citep{Giare:2023ejv}. However, there are no easily identifiable systematic uncertainties associated with the low-$\ell$ EE measurement from Planck, and it is difficult to accommodate such a high value of $\tau_{\rm reio}$ when these data are included. Nonetheless, a model that could accommodate the large-scale polarization data from Planck with a high optical depth remains a possible solution \citep[e.g.][]{Namikawa:2025doa}.

\subsubsection{Evolving dark energy}

Finally, given the implications for the post-recombination expansion history, it is worth mentioning the impact of evolving dark energy, and whether it can solve either or both apparent discrepancies. The DESI Collaboration finds a $3.1\, \sigma$ preference for evolving dark energy, the significance of which increases when other probes of late-time geometry (in this case, uncalibrated supernovae \citep{Brout:2022vxf, Rubin:2023ovl,DES:2024jxu}) are included in the analysis \citep{DESI:2024mwx, DESI:2024kob, DESI:2024aqx, DESI:2025zgx}. 

Evolving dark energy models are able to fit the feature in BAO data that, in \lcdm\, leads to a high cosmological constant — the same feature driving the matter-density deficit highlighted here — without requiring a low matter density. Exactly because evolving dark energy can allow for a higher matter density, the impact of BAO data on lensing expectations in these model spaces is modified (relative to what happens with \lcdm\ + $\mnu$) to allow for higher lensing power, reducing the CMB lensing excess somewhat. We note, however, that the $w_0w_a$ evolving dark energy model does not predict a higher lensing power when fitted to CMB data alone, compared to $\lcdm$, and therefore does not address the  $\sim2\, \sigma$ lensing excess that is found even when excluding BAO data \citep{SPT-3G:2024atg}.

\citet{DESI:2025ejh} have also shown that freeing the dark energy equation of state, within their signed neutrino mass model, results in a shift in the $\mnu$ posterior towards more positive values. This is consistent with our expectations given the higher matter density discussed in the previous paragraph. In combination with the increased uncertainty in the signed mass, once the dark energy equation of state is allowed to vary, the significance of the preference for $\mnu < 0$ is reduced. When combined with any of the above supernova datasets, however, the peak posterior shifts back towards negative values to varying degrees.

We should mention as well that in extended parameter spaces, including evolving dark energy and a free neutrino mass (in addition to other extension parameters), \citet{RoyChoudhury:2024wri} have found indications of positive neutrino mass and $A_{\rm 2pt}$ consistent with unity — but no preference for evolving dark energy, and with increased uncertainties due to the additional degeneracies. \citet{Escudero:2024uea} likewise studied a model space where both the dark energy equation of state and $\Sigma m_{\nu}$ were among the parameters allowed to vary, and found an approximately $2\, \sigma$ preference for negative signed neutrino mass by fitting a Gaussian to constraints on $\Sigma m_{\nu}$.

\section{\label{sec:conclusion} Conclusions}

We have two different suggestions of a problem in the \lcdm\  + $\mnu$ model space, given BAO and CMB data, which we have referred to as a lensing power excess and a matter-density deficit. These are not statistically overwhelming discrepancies. \citet{Craig:2024tky} quantify the first as a $2.4\, \sigma$ discrepancy, which with the addition of the SPT-3G polarization-only CMB lensing power becomes a $2.7\, \sigma$ discrepancy \citep{SPT-3G:2024atg}. Here we have quantified the matter-density deficit (relative to expectations under minimal NO) as a $2.3\, \sigma$ discrepancy. 

The focus of this paper is a description of the problem. We quantified the matter-density deficit and provided a narrative aimed at analytic understanding of the origin of cosmological constraints on $\mnu$ from current CMB and BAO data.

We first summarized an analysis by \citet{Loverde:2024nfi}. In brief, BAO data are sensitive to the combination $\omm \rs^2$. Since CMB data, in the \lcdm\ + $\mnu$ model, can be used to determine $\rs$, the combination can be used to infer $\omm$. The precision of the $\omm$ inference is then greatly enhanced by including the very tight constraint on $\thetas$ as inferred from CMB data. This is because to stay at fixed $\thetas$ a $\delta \omnu$ must be compensated by a 13 times larger change in the cosmological constant energy density. Finally, since CMB data can determine $\omega_{\rm b}$ and $\omega_{\rm c}$, one can compare with $\omm = \omnu + \omega_{\rm b} + \omega_{\rm c}$ to determine $\omega_\nu = \mnu/93.14$ eV.

Current primary CMB are sensitive to a linear combination that is approximately $\omcb - 0.17 \omnu$, owing to the lensing-induced peak smearing of CMB power spectra, and this linear combination changes to approximately $\omcb - 0.7 \omnu$ when lensing reconstruction data are included. We explain this change in sensitivity as arising due to the different scales being probed by two-point versus four-point lensing, with the latter probing smaller scales where neutrino masses have a greater effect on lensing power. We expect that inclusion of more lensing reconstruction data with more weight towards higher $L$ will increase this sensitivity. Due to the different linear combinations of $\omcb$ and $\omnu$ that BAO and CMB lensing data constrain (with BAO data constraining $\omcb+\omnu$, we also argued that the BAO and CMB lensing complement each other as probes of the neutrino mass.

We have also quantified the second of the aforementioned problems in the $\lcdm+\mnu$ model space, namely the matter-density deficit. Using a phenomenological model, and comparing to the density expected in \lcdm\ with $\mnu=0.06\, {\rm eV}$, we find a $2.3\, \sigma$ preference for a deficit, when using the Planck PR3 likelihood. This preference is reduced to $1.6\, \sigma$ when using a Planck PR4 likelihood, and further reduced to $0.9\, \sigma$ when marginalizing over $A_{\rm 2pt}$ with the PR3 likelihood. For the most constraining dataset we consider, P-ACT-b, we find a $1.9\, \sigma$ preference. These changes in significance between datasets result from shifts in the inferred $\omcb$ to lower values, reducing the difference $\omm - \omcb$.

We discussed potential solutions to the excess lensing and matter-density deficit, and how the complementary nature of the CMB lensing and BAO probes can be used to discriminate among them. We studied one of these potential solutions in particular, a model of decaying dark matter, along with free $\mnu$, and found that while such a model could eliminate the matter-density deficit, it increases the lensing excess.

As measurements continue to improve, the framework developed here may be useful in understanding how different cosmological probes contribute to constraints on the neutrino mass. 

{\small
\section*{Acknowledgements}
The authors thank Zachary Weiner, Marilena Loverde, Fei Ge, Arsalan Adil, and Uros Seljak for many helpful discussions. The authors were supported in part by DOE Office of Science award DESC0009999. LK also thanks Michael and Ester Vaida for their support via the Michael and Ester Vaida Endowed Chair in Cosmology and Astrophysics.}

\bibliography{main}
\appendix

\section{\label{app:lkl_comparison}Comparison of background-only and full likelihoods}

In the analysis presented in Section~\ref{sec:background_analysis}, we constrain the signed neutrino mass model using a set of ``background-only" likelihoods, which are meant to encapsulate the CMB constraint on the post-recombination background evolution. However, as was shown in Fig.~\ref{fig:mnu_constraints_physical}, the compression to the background-only likelihood is not lossless. In this appendix we report additional checks we performed while isolating the origin of the difference between the full and background-only likelihood constraints.

\begin{figure}[t]
    \centering
    \includegraphics{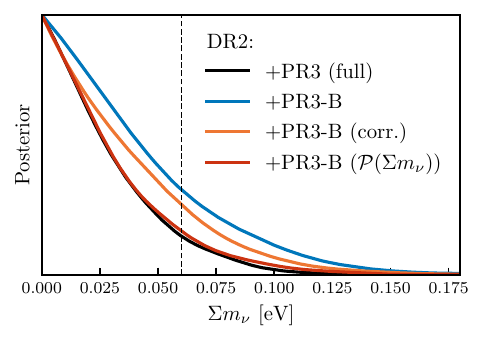}
    \caption{One dimensional marginalized posteriors for $\mnue = \mnu$, using either the full likelihood or the background-only compression with modifications, as described in the text.}
    \label{fig:appendix_plot}
\end{figure}

As discussed in Section~\ref{sec:theory}, primary CMB information is chiefly sensitive to a linear combination of $\omcb$ and $\omnu$ that is approximately $\omcb - 0.13\omnu$. This slight degeneracy results in a loss of information on $\omcb$ if $\omnu$ is simultaneously allowed to vary. As such, the compression to $\mathcal{P}(\omega_{\rm b}, \omega_{\rm c}, \thetas)$ is not the most efficient compression that one can perform; a more efficient choice would include the more tightly constrained quantity, $\omcb - 0.13\omnu$.

To assess the extent to which this inefficiency contributes to the information loss, we apply a ``correction" to the background-only likelihood. To apply the correction, we pass the quantity $\omega_{\rm c} - 0.13(\mnue/{\rm eV} - 0.06)/93.14$ to the likelihood in place of $\omega_{\rm c}$.\footnote{The chains used to construct the background-only likelihoods assume $\mnu=0.06\,{\rm eV}$, so this correction amounts to treating the $\omega_{\rm c}$ constraint in the background-only likelihood as a constraint on the more tightly constrained linear combination instead.} The resulting constraint on $\mnue$ is shown as the orange curve in Fig.~\ref{fig:appendix_plot}. While some information is recovered by changing the compression in this way, the constraint is still noticeably weaker than the constraint from the full likelihood.

We have also run a chain in which we include a one-dimensional prior on $\mnue$, which we choose to be equal to the marginalized 1D posterior on $\mnu$ coming from the full likelihood chain. This is done to rule out errors in the model implementation or the interaction of the theory code with the BAO likelihood. As expected, including this prior recovers the full likelihood result: we show this as the red curve in Fig.~\ref{fig:appendix_plot}.

We therefore conclude that information regarding $\omnu$ is lost in the compression to the background-only likelihood. As extensively discussed in Section~\ref{sec:theory}, primary CMB data constrain $\mnu$ via lensing effects, and via the constraint on the distance to last scattering. Since the latter depends only on $\thetas$ and the details of recombination, for which it is sufficient to know $\omega_{\rm b}$ and $\omega_{\rm c}$, it is completely encapsulated by the background-only compression we have used. Conversely, the lensing information in the full likelihood is compressed mainly into the value of $\omega_{\rm c}$ in the background-only likelihood, despite the fact that CMB lensing does not only depend on the matter density. We therefore suspect that the information on $\mnu$ lost during compression is almost certainly arising from the effect of lensing on the CMB primary power spectra.
\end{document}